\documentclass[aps,prb,twocolumn,superscriptaddress,
               amsmath,a4paper,floatfix]{revtex4}

\usepackage{graphicx}
\usepackage{dcolumn}
\usepackage{bm}
\usepackage{subfigure}
\usepackage{color}
\usepackage{amsmath}
\usepackage{amssymb}
\usepackage{ulem}

\graphicspath{{../}}

\begin{document}
\title{Effective Spin-1/2 Description of
   Transverse-Field-Induced Random Fields in Dipolar Spin Glasses
   with Strong Single-Ion Anisotropy}
\date{\today}
\author{S. M. A. Tabei}
\affiliation{Department of Physics and Astronomy,
University of Waterloo, Waterloo, Ontario N2L 3G1, Canada}
\author{F. Vernay}
\affiliation{Department of Physics and Astronomy,
University of Waterloo, Waterloo, Ontario N2L 3G1, Canada}
\author{M. J. P. Gingras}
\affiliation{Department of Physics and Astronomy,
University of Waterloo, Waterloo, Ontario N2L 3G1, Canada}
\affiliation{Department of Physics and Astronomy, University of Canterbury,
Private Bag 4800, Christchurch, New Zealand}
\affiliation{Canadian Institute for Advanced
 Research, 180 Dundas St. W., Toronto, Ontario, M5G 1Z8, Canada}

\begin{abstract}
We present analytical and numerical evidence for the validity of an effective
$S_{\rm eff}=\frac{1}{2}$
approach to the description of random field generation in  $S\geq1,$ and especially
in an $S=1$, dipolar spin glass models
with strong uniaxial Ising
anisotropy and subject to weak external magnetic field $B_x$ transverse to the
Ising direction.
Explicitely $B_x-$dependent random fields are shown to naturally emerge
in the effective low-energy description of a microscopic $S=1$ toy model.
We discuss our results in relation to recent theoretical studies pertaining to
the topic of $B_x-$induced random fields in the
 LiHo$_x$Y$_{1-x}$F$_4$ magnetic materials with the Ho$^{3+}$ Ising moments subject
to a transverse field.
We show that the $S_{\rm eff}=\frac{1}{2}$ approach is able to capture both
the qualitative and {\it quantitative} aspects of the
physics at small $B_x$, giving results that agree with those obtained using
conventional second order perturbation theory.

\end{abstract}

\maketitle

\section{Introduction}

In condensed matter physics systems with strongly interacting quantum mechanical
degrees of freedom, it is often a challenge to explain
physical phenomena from a truly first principle atomistic point of view.
In systems where there are high energy scales well separated from a low-energy sector,
effective low-energy theories offer the advantage of a  reformulation of
the problem with an exponentially smaller Hilbert space.
A well known and topical example where such an approach is used is
in the derivation of an effective
spin-only model starting from a Hubbard model describing electrons hopping
on a lattice. It is commonly accepted that the
low-energy magnetic excitations of a Hubbard model with a large Coulomb repulsion $U$
are easier to investigate within an effective spin Hamiltonian.
\cite{MacDonald,Sacha,Delannoy} Generally speaking, the only requirement
to be able to derive an effective model is to have a small parameter,
which is $t/U$ in the previous example, where $t$ is the nearest-neighbor
hopping constant.

In many magnetic materials, the ground state
degeneracy of the otherwise free magnetic ions can be partially lifted by
electrostatic and covalent interactions due to the surrounding atoms
$-$ the so called crystal field effect.
In a number of situations, the energy scales associated with the spin-spin interactions
are much smaller than the energy gap between the single-ion ground state and the excited
crystal field states. In such cases, one can, as a first approximation, often neglect
the high energy states and reduce the relevant Hilbert space to a much smaller
subspace of low energy states.
In this paper, we discuss the quantitative validity of
an effective low-energy theory description of a model inspired by the
phenomena displayed by the disordered LiHo$_x$Y$_{1-x}$F$_4$
magnetic material when subject to an external magnetic field $B_x$
applied perpendicular to the Ising direction of the Ho$^{3+}$ magnetic moments.

The LiHo$_x$Y$_{1-x}$F$_4$ magnetic material
exhibits many interesting magnetic
behaviors.\cite{Reich,Wu,Ghosh-Science,Ghosh-Nature,Ronnow-Science,Silevitch-Nature,Silevitch-PRL}
The magnetic properties of LiHo$_x$Y$_{1-x}$F$_4$ are due to the Ho$^{3+}$ ions.
The single-ion ground state of Ho$^{3+}$
is a doublet, while the first excited state is at
$\sim 11$ K above the ground state.\cite{Hansen,Ronnow-PRB}
The most relevant
 interactions between the magnetic Ho$^{3+}$ ions are
magnetic dipole-dipole interactions.\cite{Chakraborty}
Since the maximum strength of the dipolar interactions is for
nearest neighbor separation and is approximatly 0.31 K,
  collective behavior in this material occurs at temperatures
less than O(1 K) where
only the ground doublet is significantly thermally populated.
Consequently, the cooperative phenomena
and the low temperature properties of this material
in zero applied magnetic field
should be well captured by an effective model with
 spin-$\frac{1}{2}$ degrees of freedom.\cite{Chakraborty,Bitko}
For example, in zero applied magnetic field,
the system can  be recast as a diluted dipolar Ising model with the low-temperature phase
being either a ferromagnet or a spin glass depending on the concentration $x$
of magnetic ions.\cite{Reich,antiglass}
On the other hand, for $x=1$ and with a magnetic field $B_x$
applied perpendicular to the crystallographic Ising $c-$axis direction,
LiHoF$_4$ has  been advocated as one of the rare physical realization~\cite{Bitko}
of the transverse field Ising model (TFIM).\cite{deGennes,Pfeuty,Sachdev,TFIM-book,Stasiak}
Yet, it is only relatively recently that a somewhat rigorous justification of
a TFIM description of LiHoF$_4$ in nonzero $B_x$ has been put forward.\cite{Chakraborty}
However, over the past twenty years, and until very recently,
several experimental studies had found the behavior of
LiHo$_x$Y$_{1-x}$F$_4$ ($x<1$, $B_x\ne 0$) paradoxical, as we now discuss.

One may have naively expected that the application of a transverse
magnetic field in
LiHo$_x$Y$_{1-x}$F$_4$ would allow to explore the physics of the TFIM in either
a diluted ferromagnet or a spin glass, depending on the concentration $x$.
However, the situation for $x<1$ is quite a bit more complicated.\cite{Wu}
For example,
 for $B_x=0$,
  LiHo$_{0.167}$Y$_{0.833}$F$_4$ displays a conventional
  spin glass phase transition~\cite{Wu,no-sg}
   with a nonlinear magnetic susceptibility, $\chi_3$,
diverging at the spin glass transition temperature, $T_g$,
 as $\chi_3(T)\propto (T-T_g)^{-\gamma}$ as in
ordinary spin glass materials.\cite{Mydosh}
However, as $B_x$ is increased from zero,
$\chi_3(T)$ becomes steadily less singular,
and there appears to be no $B_x-$induced quantum critical
phase transition between a paramagnet and a spin glass state.\cite{Wu}
This puzzling experimental behavior had been tentatively interpreted
as due to a 1$^{\rm st}$ order transition near the $T=0$ quantum
phase transition.\cite{Wu,Grempel}
However, very recent and independent theoretical
investigations~\cite{Schechter-PRL2,Tabei,Schechter-Osaka}
have instead proposed that the microscopic origin of the ``quenching''
of the paramagnetic to
spin glass transition as $B_x$ is turned on
is due to the generation of random fields that destroy
the spin glass phase.

The authors of Ref.~[\onlinecite{Tabei}] used
an effective $S_{\rm eff}=1/2$ theory, very similar to the one
developped for pure LiHoF$_4$ \cite{Chakraborty} to expose how
random fields develop in a {\it microscopic model} of
LiHo$_x$Y$_{1-x}$F$_4$ in nonzero $B_x$.
In particular, Ref.~[\onlinecite{Tabei}] showed how the nonlinear susceptibility
$\chi_3$ becomes progressively less singular as $B_x$ is increased.
Also motivated by the phenomena displayed by  LiHo$_x$Y$_{1-x}$F$_4$,
Schechter and collaborators
\cite{Schechter-PRL2,Schechter-Osaka,Schechter-3}
also recently investigated
in a series of papers the general phenomenology of induced random fields in
 LiHo$_x$Y$_{1-x}$F$_4$.
To do so, they considered in Refs.~[\onlinecite{Schechter-PRL2,Schechter-Osaka}]
an easy-axis  spin-$S$ ($S\geq 1$)
dipolar spin glass toy model Hamiltonian, ${\cal H}$, in presence of a nonzero $B_x$.
By using second order perturbation theory, invoking the scaling droplet
picture of Fisher and Huse for spin glasses,~\cite{Fisher}
and using an Imry-Ma type argument,\cite{Imry}
Schechter {\it et al.}~\cite{Schechter-PRL2,Schechter-Osaka}
calculated the
finite energy $\delta E$
required to flip the spins within a spin glass droplet,
finding a limit on how large the spin glass correlation length $\xi$ can grow to
as the system is cooled from the paramagnetic phase.
The behavior of the system, and the corresponding $\delta E$, is found to be analogous to that
of a spin glass in a random magnetic field which, according to the droplet model,
does not show a spin glass transition in nonzero field.\cite{Young-RFSG}
As a result, Refs.~[\onlinecite{Schechter-PRL2,Schechter-Osaka}] argue
that no spin glass transition can occur in a dipolar spin glass where
random off-diagonal dipolar interactions and an applied transverse magnetic field
are simultaneously at play.

On one hand, the results of both Refs.~[\onlinecite{Schechter-PRL2,Schechter-Osaka}] and
Ref.~[\onlinecite{Tabei}] derive from the notion that, the
applied transverse field generates, through the off-diagonal part
of the dipolar interactions,
which couple the Ising $\hat z$ component with the perpendicular
$\hat x$ and $\hat y$ components,
 some effective
random fields.
However, it has so far not been clarified to what extent the
random fields are quantitatively
equivalent or only qualititatively related in those two sets of works.
In their studies,  the authors of Refs.~[\onlinecite{Schechter-PRL2,Schechter-Osaka}]
argued,  correctly, that considerations of a model with large spin ($S\ge 1$) is
crucial to understand the weak field response of the spin
glass phase in either their toy model ${\cal H}$  or in LiHo$_x$Y$_{1-x}$F$_4$.
Exact diagonalization results of an $S=1$ dipolar spin glass model with
easy-axis anisotropy provided further quantitative support to the
theoretical arguments as to the scaling behavior of
$\delta E$ with both $B_x$ and the number of spins  in the
system. \cite{Schechter-PRL2,Schechter-Osaka}
At the same time, their results
 from similar calculations~\cite{Schechter-Osaka} for
an effective anisotropic spin-$\frac{1}{2}$ dipolar Ising model in a transverse field, but
with the off-diagonal dipolar interactions rescaled compared to the longitudinal
Ising coupling,\cite{Tabei}  did not conform with those obtained for the
``bare'' (high-energy) anisotropic $S=1$ model. \cite{Schechter-PRL2,Schechter-Osaka}
Partially on the basis of those results,
and seemingly confirming a previous argument~\cite{Schechter-PRL2},
Ref.~[\onlinecite{Schechter-Osaka}],
 concludes
 that an effective spin-$\frac{1}{2}$ model, such as that used in
Ref.~[\onlinecite{Tabei}], is not sufficient to capture the physics in the small
$B_x$ regime compared to the ``bare'' microscopic (large-spin) anisotropic
dipolar spin glass model ${\cal H}$.
The question of the usefulness
 of an effective spin-$\frac{1}{2}$ model to describe
random field phenomena in the dilute ferromagnetic regime of
LiHo$_x$Y$_{1-x}$F$_4$ \cite{Tabei,Silevitch-Nature,Brooke-thesis}
has also been recently raised.\cite{Schechter-3}

Considering a perspective beyond the specific problematic
of LiHo$_x$Y$_{1-x}$F$_4$, one could interpret the conclusion of
Refs.~[\onlinecite{Schechter-PRL2,Schechter-Osaka,Schechter-3}]
 regarding the
inadequacies of
an effective spin-$\frac{1}{2}$ model to describe LiHo$_x$Y$_{1-x}$F$_4$ in $B_x\ne 0$ as a
counter example
of the precise
 quantitative
usefulness of effective low-energy
theories for
quantum $N$-body systems. It is therefore
useful
to investigate with some scrutiny the mathematical
justification for an effective spin-$\frac{1}{2}$ model for LiHo$_x$Y$_{1-x}$F$_4$
with $B_x\ne 0$.
This is the purpose of the present paper.
More specifically, the question that we ask here is: to what extent are
the explicitely manifest random fields derived in an effective low-energy
theory, such as in Ref.~[\onlinecite{Tabei}],
related to the random field like effects at play in perturbation
theories, such as used in Refs.~[\onlinecite{Schechter-PRL2,Schechter-Osaka}]?
Below we show, via a derivation of an effective low-energy
$S_{\rm eff}=\frac{1}{2}$ Hamiltonian for anisotropic dipolar glasses, that effective random
longitudinal fields emerge naturally in the  $S_{\rm eff}=\frac{1}{2}$ model.
On the basis of analytical calculations and exact
diagonalizations, we highlight the fact that
an $S_{\rm eff}=\frac{1}{2}$ Hamiltonian properly
derived from an  $S=1$ high-energy toy model $\cal{H}$, such as the one
proposed in Refs.~[\onlinecite{Schechter-PRL2,Schechter-Osaka}]
(see Eq. (1) in Section ~\ref{Hamiltonian}),
is a quantitatively valid and controlled approach to this problem.

The paper is organized as follows.
We first discuss in Section~\ref{Hamiltonian} an
anisotropic spin-$S$ dipolar Hamiltonian as a simplified model displaying the
key physics of the LiHo$_x$Y$_{1-x}$F$_4$ material in a transverse field and show
in Section~\ref{Effective-H}
how to derive from it an effective $S_{\rm eff}=\frac{1}{2}$ Hamiltonian to lowest order.
We present in Section~\ref{Numerical}
results from exact diagonalization calculations that compare
the $S=1$ and the $S_{\rm eff}=\frac{1}{2}$ models and which directly
confirm the quantitative validity of the effective Hamiltonian approach.
Section~\ref{Conclusion} concludes the paper.


\section{Anisotropic Spin Hamiltonian}
\label{Hamiltonian}

The Ho$^{3+}$ ion is characterized by a very large hyperfine interaction between the
electronic and nuclear moments and the effects of this strong interaction
plays an important role in a number of Ho$^{3+}-$based magnetic
materials.\cite{Ronnow-Science,Ronnow-PRB,Bitko,Ramirez-Jensen,Bramwell}
In particular, in LiHoF$_4$, it leads to a significant increase of the zero
temperature critical transverse field for the dipolar ferromagnet to quantum
paramagnet transition.\cite{Chakraborty,Bitko} It also plays an important
role in setting the relevant critical transverse magnetic field scale
in the dilute LiHo$_{x}$Y$_{1-x}$F$_4$.\cite{Schechter-PRL1}
In this paper, however, we are specifically
interested in the general phenomenology of
random fields along the Ising spin directions generated
by {\it small} applied transverse field rather than obtaining a
precise quantitative description of LiHo$_x$Y$_{1-x}$F$_4$.
In this specific context, we therefore neglect
the role of hyperfine interactions.
Also neglecting the hyperfine interactions,
Schechter {\it et al.}~\cite{Schechter-Osaka,Schechter-PRL2}
proposed a generic anisotropic spin-$S$ toy model
Hamiltonian with long-range dipolar interactions
\begin{eqnarray}
{\cal{H}}&=&- D \sum_i [(S_i^z)^2 - S^2]\nonumber\\
&-&\sum_{i \neq j}
\left[\frac{1}{2}V_{ij}^{zz}S_i^z  S_j^z+V_{ij}^{zx}
S_i^z  S_j^x\right] -B_x\sum_i S_i^x  ~ .
\label{anisH}
\end{eqnarray}
This Hamiltonian is a simplified model that preserves the basic characteristics of the proposed
 microscopic Hamiltonian~\cite{Chakraborty,Tabei} for LiHo$_x$Y$_{1-x}$F$_4$.
  In the absence of an external field,
individual Ho$^{3+}$ spins have an Ising like ground state doublet
with a large energy gap between the next excited state and the ground doublet.
Also, for $S=1$, the excited state of model in Eq.~(\ref{anisH})
is a singlet, as for Ho$^{3+}$ in
LiHo$_{x}$Y$_{1-x}$F$_4$.\cite{Hansen,Ronnow-PRB}
 Here, $i,j$ are the positions of the randomly
positioned magnetic moments.
$V_{ij}^{\mu \nu}$ denotes the random long-range dipolar interaction between the spins,
 where
$V_{ij}^{zz}$ stands for the Ising interaction and $V_{ij}^{zx}$ stands for the off-diagonal
interaction ($V_{ij}^{\mu\nu}=V_{ij}^{\nu\mu}$ for dipolar interactions).
$D>0$ is the anisotropy constant mimicking the crystal field.
For $B_x=0$, the ground state (GS) of a single spin is doubly
degenerate with $S^z=\pm S$~.  The corresponding states of the doublet
are denoted $|S\rangle$ and $|-S\rangle$.
The first excited states have
 $S^z=\pm (S-1)$ and energy $\Omega_0\equiv(2S-1)D$,
with the corresponding states denoted as
$|\pm (S-1)\rangle$.
Ignoring momentarily the $V_{ij}^{\mu\nu}$ interactions,
the Zeeman term, $-B_x\sum_i S_i^x$,
lifts the GS degeneracy of the
 $|\pm S\rangle$ ground doublet, resulting in two new lowest energy states,
$|\alpha(B_x)\rangle$ and $|\beta(B_x)\rangle$,
with corresponding energies $E_{\alpha}(B_x)$ and $E_{\beta}(B_x)$, and
 with an energy gap
\begin{equation}
 \Delta(B_x)=E_{\alpha}(B_x)-E_{\beta}(B_x)
 \label{Delta}
\end{equation}
between them.
For $B_x\ll\Omega_0$, to leading order in perturbation theory,
the gap $\Delta(B_x)$ is proportional to $(B_x)^{2S}$.\cite{B2s}

Invoking  the spin glass droplet scaling picture
 of Fisher and Huse,\cite{Fisher} and using an Imry-Ma~\cite{Imry} type argument,
one can calculate the energy required to
flip a spin glass droplet of size $L$ containing $N\sim L^d$ spins,
with $d$ the number of space dimensions (here $d=3$).
This energy cost is due to the perturbative quantum
${\cal H}_{\perp} \equiv -\sum_{i \neq j}V_{ij}^{zx} S_i^z  S_j^x
-B_x\sum_i S_i^x $ which term does not commute with the
the classical
${\cal H}_{\parallel}=- D \sum_i [(S_i^z)^2 - S^2]
-\frac{1}{2}\sum_{i \neq j}V_{ij}^{zz} S_i^z  S_j^z$ term.
Considering first only ${\cal H}_{\parallel}$, and taking the droplet picture of
only two distinct ground states,\cite{Fisher}
~$|\Phi_{S}\rangle$ and~$|\widetilde{\Phi}_{S}\rangle$ denote the
collective (doubly-degenerate) Ising spin glass ground states  of the system.
These two ground states are related by the global
$S_i^z\rightarrow -S_i^z$  symmetry,
where each spin is either in its $|+S\rangle$ state or its $|-S\rangle$ state.
 As discussed in Refs.~[\onlinecite{Schechter-PRL2,Schechter-Osaka}], nonzero
${\cal H}_{\perp}$
lifts the ground state degeneracy, as we now
review in order to make contact with the results presented below in Sections
\ref{Effective-H}
and \ref{Numerical}.

The lowest energy excited states (above the  otherwise two degenerate
~$|\Phi_{S}\rangle$ and~$|\widetilde{\Phi}_{S}\rangle$ ground states)
are $|\phi^k_{(S-1)}\rangle$ and $|\widetilde{\phi}^k_{(S-1)}\rangle$ states,
in which the $k$'th spin
has its $S^z$ quantum value changed from $+S$ to $+(S-1)$ or from $-S$ to $-(S-1)$.
 Using standard second order degenerate perturbation theory,\cite{pert}
 and considering only excitations to
the (intermediate excited) $|\phi_{(S-1)}\rangle$ and $|\widetilde{\phi}^k_{(S-1)}\rangle$
states, the fluctuation-induced energy difference
between ~$|\Phi_{S}\rangle$ and~$|\widetilde{\Phi}_{S}\rangle$ is
\begin{eqnarray}
\delta E&=&
\sqrt{ \left(
H_{\Phi_S,\Phi_S}  -  H_{{\widetilde{\Phi}_{S}},{\widetilde{\Phi}_{S}}}
\right)^2
+4|H_{\Phi_S,\widetilde{\Phi}_S}|^2 }
\label{deltaE1}
\end{eqnarray}
where
\begin{eqnarray*}
H_{\Phi_S,\Phi_S}&=&
-\frac{1}{\Omega_0}\sum_k\left|\left\langle \Phi_S\left|H_{\perp}\right|\phi^k_{ (S-1)}\right\rangle\right|^2 ~, \\
H_{\widetilde{\Phi}_S,\widetilde{\Phi}_S}
&=&-\frac{1}{\Omega_0}\sum_k\left|\left\langle \widetilde{\Phi}_S\left|H_{\perp}\right|
\widetilde{\phi}^k_{(S-1)}\right\rangle\right|^2 ~,
\end{eqnarray*}
and
\begin{eqnarray*}
H_{\Phi_S,\widetilde{\Phi}_S}=
& - & \frac{1}{\Omega_0}\sum_k\left\langle \Phi_S\left|H_{\perp}\right|\phi^k_{(S-1)}\right\rangle
\left\langle\phi^k_{(S-1)}|H_{\perp}|\widetilde{\Phi}_S\right\rangle    \\
& + & \left\langle \Phi_S\left|H_{\perp}\right|\widetilde{\phi}^k_{(S-1)}\right\rangle
\left\langle\widetilde{\phi}^k_{(S-1)}|H_{\perp}|\widetilde{\Phi}_S\right\rangle ~ .
\end{eqnarray*}
where we have taken the ground state energy to be zero.
Since $\langle \Phi_S\left|H_{\perp}\right|\widetilde{\phi}^k_{(S-1)}\rangle=
\langle\widetilde{\Phi_S}\left|H_{\perp}\right|\phi^k_{(S-1)}\rangle=0$, we have $H_{\Phi_S,\widetilde{\Phi}_S}=0$~
Subtracting
$H_{\widetilde{\Phi}_S,\widetilde{\Phi}_S}$ from
$H_{\Phi_S,\Phi_S}$,
only the odd terms in $B_x$ remain, with the even terms in $B_x$ cancelling  each other out.
Finally, to lowest order in $B_x$, we get
\begin{eqnarray}
\delta E=2S\frac{B_x}{\Omega_0}\sum_{i\neq j}V_{ij}^{xz}\left\langle\Phi_s\vert S_i^z\vert\Phi_s
 \right\rangle~.
 \label{deltaEAn}
\end{eqnarray}
 Taking the largest $V_{ij}^{xz}$ with a typical value $V_\perp$,
the typical energy gained by flipping a droplet of $N\sim L^d$ spins is,
 to leading order in $B_x$,
\begin{equation}
\langle \vert \delta E \vert \rangle \propto \frac{S^2 B_x V_\perp \sqrt{N}}{\Omega_0},
\label{deq}
\end{equation}
indicating that the total energy gain increases with $B_x$ linearly to leading order,
as first found in Refs.~[\onlinecite{Schechter-Osaka,Schechter-PRL2}].

This decrease in energy is to be
compared with the energy cost due to the formation of a spin glass droplet.\cite{Fisher}
This energy cost scales with the  linear size $L$ of the droplet, $L=N^{1/3}$,
 as $\approx S^2V_{\parallel} L^{\theta_d}$,
where $V_\parallel$ is the typical value of the largest $V_{ij}^{zz}$,
which one typically expects to be of the same order as $V_\perp$.
Comparing the energy gain $\langle \vert \delta E\vert\rangle$
of Eq.~(\ref{deq}) with the energy cost for droplet formation,
Refs.~[\onlinecite{Schechter-PRL2,Schechter-Osaka}]
find a finite correlation length $\xi$,
identified with $L$, which, for \textit{small} $B_x$,
scales as
\begin{equation}
\label{corrlength}
\xi\approx \left(
\frac{\Omega_0 V_\parallel}{B_x V_\perp}
\right )^{\frac{1}{3/2-\theta_d}} .
\end{equation}
Based on an argument by Fisher and Huse,\cite{Fisher}
$\ \theta_d\leq\ $ $\ (d-1)/2$, or $\theta_d<3/2$ here.
Hence, turning on $B_x$ leads to a reduction of the correlation length $\xi(B_x)$,
inhibiting its divergence as occurs when $B_x=0$.
In other words, the presence of the applied transverse $B_x$ leads, via the presence
of the off-diagonal $V_{ij}^{xz}$ spin-spin interactions, to a destruction of the spin glass
phase with a typical spin glass correlation length $\xi$ decreasing as $B_x$ increases.
As argued in Refs.~[\onlinecite{Schechter-PRL2,Schechter-Osaka}],
this is the mechanism via which the non-linear magnetic susceptibility $\chi_3$
no longer diverges in LiHo$_x$Y$_{1-x}$F$_4$ as $B_x$ is increased from zero.\cite{Wu,diff-chi3}


\section{Effective spin-$\frac{1}{2}$ Description}
\label{Effective-H}

In the previous section we reviewed the arguments
of Refs.~[\onlinecite{Schechter-PRL2,Schechter-Osaka}]
which lead to the key result of Eq.~(\ref{deltaEAn}).
We now proceed to show that a reformulation of
the microscopic spin Hamiltonian, Eq.~(\ref{anisH}), in terms of an effective
$S_{\rm eff}=\frac{1}{2}$ model, leads identically to Eq.~(\ref{deltaEAn})
in the limit of small $B_x/D$.

Firstly, we focus on a situation where the temperature considered is low
 compared to $\Omega_0$, and project the spin $S$
operators onto the two-dimensional subspace formed by  the
 two lowest energy eigenstates, $|\alpha(B_x)\rangle$ and $|\beta(B_x)\rangle$.
Following Refs.~[\onlinecite{Chakraborty,Tabei}],
we define an Ising subspace, $|\! \uparrow\rangle$ and $| \! \downarrow\rangle$,
  by performing a rotation
\begin{eqnarray}
\vert \! \uparrow\rangle   & = &
\frac{1}{\sqrt{2}}(|\alpha\rangle + \exp(i\theta)|\beta\rangle)\nonumber\\
\vert \! \downarrow\rangle & = &
\frac{1}{\sqrt{2}}(|\alpha\rangle - \exp(i\theta)|\beta\rangle).
\label{updown}
\end{eqnarray}
The phase $\theta$ is chosen such that the matrix elements of the
operator $S^{z}$ within the new (Ising) subspace are real and diagonal.
In this case, we can define $S^{z}_{i}=C_{zz}\sigma^{z}_{i}$.
This allows us to recast ${\cal H}$ in Eq.~(\ref{anisH}) in terms of an effective
spin$-\frac{1}{2}$ Hamiltonian,
${\cal H}_{\rm eff}$,
 that involves the $\sigma^{\mu}$ Pauli matrices.\cite{Chakraborty}
In this projected subspace, a transverse field $\Gamma=\frac{1}{2}\Delta(B_x)$  acts on
the effective $\sigma_i^x$ spin.
The projected  $S_i^\mu$ ($\mu=x,y,z$) operator may be written as:
\begin{eqnarray}
S_i^\mu=\sum_\nu C_{\mu\nu}(B_x)\sigma_i^\nu +C_{\mu 0 }(B_x) \openone.
\label{Cmunu}
\end{eqnarray}
The $C_{\mu \nu}$ and $\Delta$ dependence on $B_x$
can be obtained by exact diagonalization~\cite{Chakraborty,Tabei}
of the non-interacting part of ${\cal H}$ (i.e. $V_{ij}^{\mu\nu}=0$) in Eq.~(\ref{anisH}).

\begin{figure}[]
\includegraphics*[width=\columnwidth,angle=0]{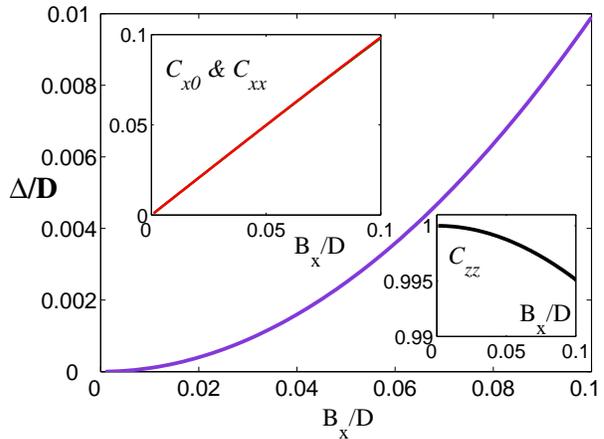}
\caption[figure1]{\label{S1coeff} (Color Online)
Evolution of $\Delta$, $C_{zz}$,
$C_{x0}$, and $C_{xx}$ as a function of the external transverse field $B_x$
for $S=1$.
}
\end{figure}
For zero transverse field, $B_x=0$, the only nonzero $C_{\mu \nu}$ coefficient is
$C_{zz}(0)=S$, giving a ``\textit{classical}'' (effective)
low-energy dipolar Ising model
\begin{eqnarray}
{\cal H}_{\rm Ising}=-\frac{1}{2}S^2\sum_{i \neq j} V_{ij}^{zz}
\sigma_i^z \sigma_j^z .
 \label{ising}
\end{eqnarray}
Turning on $B_x$, the coefficients $C_{x0}$ and $C_{xx}$ increase with $B_x$, while $C_{zz}$
shows a slight decrease with increasing $B_x$,
 as  shown in Fig.~\ref{S1coeff}.
Thus, by substituting $S_i^z$ with $C_{zz}(B_x)\sigma_i^z$
 and $S_i^x$ with $C_{xx}(B_x)\sigma_i^x+C_{x0}(B_x) \openone$
in Eq.~(\ref{anisH}),
the effective spin-$\frac{1}{2}$ Hamiltonian is
\begin{eqnarray}
\label{Heff}
{\cal H}_{\rm eff}&=&-\frac{1}{2}C_{zz}^2(B_x)\sum_{i \neq j} V_{ij}^{zz}
\sigma_i^z \sigma_j^z\nonumber\\
&& - C_{zz}(B_x)\left\{ C_{xx}(B_x)\sum_{i \neq j} V_{ij}^{zx}
\sigma_i^z \sigma_j^x\right.\\
&& -\left. C_{x0}(B_x)\sum_{i \neq j}
 V_{ij}^{zx}\sigma_i^z\right\}
-\frac{1}{2}\Delta (B_x)\sum_{i}\sigma_i^x~.\nonumber
\end{eqnarray}
As  can be seen, the projection of the $V_{ij}^{zx}
S_i^z  S_j^x$ term in Eq.~(\ref{anisH}) results in an induced random bilinear coupling,
 $\propto\sigma_i^z\sigma_j^x$,
and a longitudinal random field interaction,
  $\propto\sigma_i^z$,  for $B_x\neq 0$.
For low enough transverse field $B_x$, the Ising dipolar interaction ($\propto V_{ij}^{zz})$
is the dominant term.

Having derived the effective Hamiltonian, we
now repeat the calculation of $\delta E$
within this effective $S_{\rm eff}=\frac{1}{2}$ framework by again bringing in the spin glass
droplet picture~\cite{Fisher}.
 For $B_x=0$, we denote
$\vert\psi\rangle$ the ground state of the $S_{\rm eff}=\frac{1}{2}$ system
where  $\vert\psi\rangle$ is a specific realization of
the $\uparrow$ and $\downarrow$ (effective) Ising spins configuration.\cite{up}
For $B_x=0$, because of time reversal symmetry,
the time reversed state $\vert\widetilde{\psi}\rangle$,
 which is obtained by flipping all the spins of
$\vert\psi\rangle$, is a ground state of the system as well,
giving a ground state
 doublet in the  ``effective spin'' droplet picture.
Carrying on a similar discussion as in the previous section and as in
Refs.~[\onlinecite{Schechter-Osaka,Schechter-PRL2}],
 at low enough $B_x$ within a droplet picture, the symmetry is broken
due the presence of the induced random fields in Eq.~(\ref{Heff}).
The energy cost to flip the spins over a droplet is,
\begin{eqnarray*}
 \delta E&\equiv & \langle\widetilde{\psi}\vert H_{\rm{eff}}\vert\widetilde{\psi}
 \rangle-\left\langle\psi\vert H_{\rm{eff}}\vert\psi\right\rangle
\end{eqnarray*}
which, to lowest order in $B_x$, gives
\begin{eqnarray}
 \delta E &\approx & 2C_{zz}C_{x0}\sum_{i\neq j}V_{ij}^{xz}\left\langle\psi\vert \sigma_i^z\vert\psi
 \right\rangle~.
\label{deltaE_Heff}
\end{eqnarray}
Although we have an exact analytical
expression for the $C_{\mu\nu}$ coefficients as a function of $B_x$
(which is available for $S\le 3/2$),
 in order to compare with Eq.~(\ref{deltaEAn}) above
and with Refs.~[\onlinecite{Schechter-Osaka,Schechter-PRL2}],
we consider the $B_x$ dependence of the $C_{\mu\nu}$ to leading order in
$B_x/D$.
 Using standard degenerate perturbation theory, for $S>1$,\cite{S}
the $|\! \uparrow\rangle$ and $|\! \downarrow\rangle$ defined in Eq.~(\ref{updown}) are,
up to second order in $B_x$, given by
\begin{eqnarray}
|\! \uparrow\rangle&=&\left(1-\frac{B_x^2}{4\Omega_0^2}S \right) |S\rangle +
\frac{B_x}{\Omega_0}\sqrt{\frac{S}{2}}|S-1\rangle\nonumber\\
|\! \downarrow\rangle&=&\left(1-\frac{B_x^2}{4\Omega_0^2}S \right) |-S\rangle +
\frac{B_x}{\Omega_0}\sqrt{\frac{S}{2}}|-S+1\rangle \; , \nonumber \\
\label{updownhalf}
\end{eqnarray}
recalling that $\Omega_0=(2s-1)D$.
Returning to Eq.~(\ref{Cmunu}), via which
the $C_{\mu\nu}$ are obtained,
\textit{e.g.},
$C_{zz}= \frac{1}{2}(\langle\uparrow \! |S^z| \! \uparrow\rangle
     - \langle\downarrow \! |S^z| \! \downarrow\rangle)$
and
$C_{x0}=\frac{1}{2}(\langle\uparrow \! |S^x| \! \uparrow\rangle
       + \langle\downarrow \! |S^x|\! \downarrow\rangle)$,
we use Eq.~(\ref{updownhalf})
to find $C_{zz}\approx S(1-\frac{B_x^2}{2\Omega_0^2})$,
$C_{x0}\approx SB_x/\Omega_0 $, $C_{xx}\propto (B_x)^{2S-1}$
($C_{xx}\approx B_x/\Omega_0$ for $S=1$),
while $\Delta\propto
(B_x)^{2S}$.\cite{B2s}
Substituting those $B_x$ dependencies back in Eq.~(\ref{deltaE_Heff}),
the dependence of the energy cost $\delta E$ is, to lowest order in $B_x$,
\begin{eqnarray}
 \delta E \approx  2S^2\frac{B_x}{\Omega_0}\sum_{i\neq j}V_{ij}^{xz}\left\langle\psi\vert \sigma_i^z\vert\psi
 \right\rangle~.
\label{cost}
\end{eqnarray}
As we can see, the energy cost obtained in the $S_{\rm eff}=\frac{1}{2}$ picture
 is identical to the energy cost given by Eq.~(\ref{deltaEAn}) obtained via second
order perturbation theory and previously  reported in
Refs.~[\onlinecite{Schechter-PRL2,Schechter-Osaka}].
Thus, Eq.~(\ref{cost}) leads to the same RMS energy
cost for flipping a droplet, given by Eq.~(\ref{deq}),
and the same $B_x$ dependence of the spin glass correlation length $\xi$ in
Eq.~(\ref{corrlength}).
Hence, we have shown that
a formally
 derived effective
$S_{\rm eff}=1/2$
Hamiltonian
 does capture quantitatively the low energy physics of the full $S$
 Hamiltonian at low
transverse fields. While the argument above was constructed for the toy model of
Eq.~(\ref{anisH}), one could proceed identically
for the full blown microscopic Hamiltonian of LiHo$_x$Y$_{1-x}$F$_4$.
Indeed, this is what is the underlying program of Ref.~[\onlinecite{Tabei}].

\section{Numerical Results}
\label{Numerical}

In the same spirit as Ref.~[\onlinecite{Schechter-PRL2,Schechter-Osaka}],
in order to investigate
to what extent our proposed low energy effective spin-$\frac{1}{2}$ model
is a good description of the full anisotropic Hamiltonian~(\ref{anisH}),
and to determine the range of transverse field
over which the above analytical small $B_x$ field results is valid,
we have
performed numerical calculations
to backup our perturbative approach.
In this section we present results from
exact diagonalizations on finite-size clusters
with open boundary conditions~\cite{boundary}.
 In order to compare the present
approach with the previous investigations done by Schechter
{\it et al.},\cite{Schechter-Osaka,Schechter-PRL2}
we work at the same constant dipole concentration $x=18.75\%$.

LiHoF$_4$ is a compound with space-group
$C^{6}_{4h}\left(I4_{1}/a \right)$ with lattice parameters
$a=b=5.175\AA$, $c=10.75\AA$, and has
4 holmium ions per unit cell positioned at
$(0,0,1/2)$, $(0,1/2,3/4)$, $(1/2,1/2,0)$ and $(1/2,0,1/4)$.\cite{Mennenga}
For LiHo$_{x}$Y$_{1-x}$F$_4$, a dilution of $x=18.75\%$
is realized by distributing randomly
$N$ magnetic moments (holmium, Ho$^{3+}$ ions) in a sample of
$\frac{16\times N}{3}$ possible sites. We have chosen samples of size
$(2a,2b,c\times\frac{N}{3})$, where $N$ is a multiple of 3. Thus,
changing the number $N$ of magnetic ions means changing the size of the sample
in the z-direction in order to keep a constant dilution.

In Eq.~(\ref{anisH}), the dipolar interaction is written as $V_{ij}^{\alpha\beta}$,
which takes, with the negative coefficient convention used in Eq.~(\ref{anisH}),  the
explicit form:
\begin{equation}
V_{ij}^{\alpha\beta}=\frac{\mu_{B}^2}{r_{ij}^3}
\left[
\frac{3 r_{ij}^{\alpha} r_{ij}^{\beta}}{r_{ij}^2}
-
\delta_{\alpha\beta}
\right],
\end{equation}
where $r_{ij}$ is the distance between the ions at positions $i$ and $j$, and
$\alpha,\beta=x,y,z$. The dipolar interaction $V^{zz}$ is of the order
$\frac{\mu_B^2}{a^3}\approx 4.49\times 10^{-3}$ K, whereas the on-site
anisotropy is taken as
$D=10$ K. In the following,
 we investigate the behavior of the
gap $\delta E$ between the ground-state and first excited-state as a function
of the applied transverse field $B_x$. Since we are mainly interested in
checking the relations (\ref{deltaEAn}), (\ref{deq})
 and (\ref{cost}),
we present our results in terms of renormalized parameters
$(\frac{\delta E}{D\sqrt{N}},\frac{B_x}{D})$.

\begin{figure}[]
\includegraphics*[width=\columnwidth,angle=0]{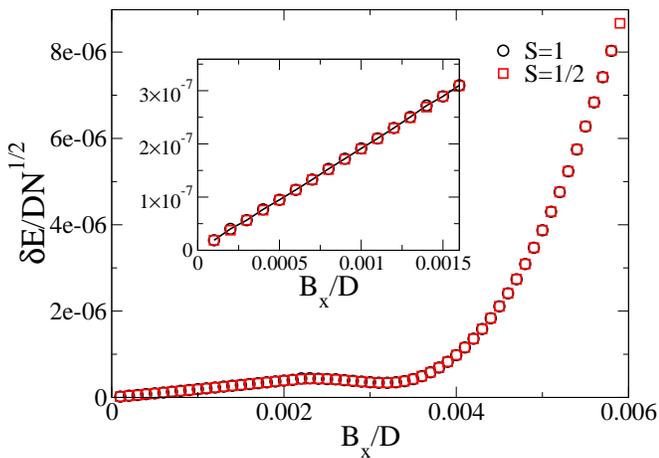}
\caption[figure2]{\label{compare} (Color Online)
Comparison between the $S=1$ and $S_{\rm eff}=1/2$
models for a given sample (e.g. realization of disorder) of $N=9$ spins:
 gap $\delta E/(D\sqrt{N})$ as a function of the transverse field $B_x/D$.}
\end{figure}

To perform a first check of the validity of our approach, we choose a small
cluster with a fixed random distribution of $N=9$ spins and compute the
renormalized gap $\delta E/(D\sqrt{N})$ for both models
({\it i.e.} $S=1$, Eq.~(\ref{anisH}) and $S_{\rm eff}=\frac{1}{2}$, Eq.~(\ref{Heff}))
as a function of the reduced transverse
magnetic field $B_x/D$. The results are shown in Fig.~\ref{compare}.
In  zero transverse field the ground-state
is degenerate and its energetics is governed by the Ising
interaction $V^{zz}$.
The application of a small transverse field $B_x$
lifts the degeneracy, with the splitting
between the ground-state and the first excited state corresponding to
the state with spins flipped.
In that regime the most important interaction remains
$V^{zz}$ and the gap $\delta E$ is found to be proportional to $B_{x}/D$
(inset of Fig.~\ref{compare}),
as suggested by the arguments leading to Eqs.~(\ref{deltaEAn}) and (\ref{cost}).
By turning on $B_x/D$ to larger values, the transverse field eventually
becomes stronger than the dipolar interactions.
At that point, the perturbative low $B_x$ regime~\cite{droplet}
 is no longer valid and the gap $\delta E$ is no longer proportional to $B_x$.
However,
Fig.~\ref{compare} shows that,
even for high transverse fields,
we observe a good agreement between the $S=1$ and the effective
$S_{\rm eff}=\frac{1}{2}$ description.

\begin{figure}[]
\includegraphics*[width=\columnwidth,angle=0]{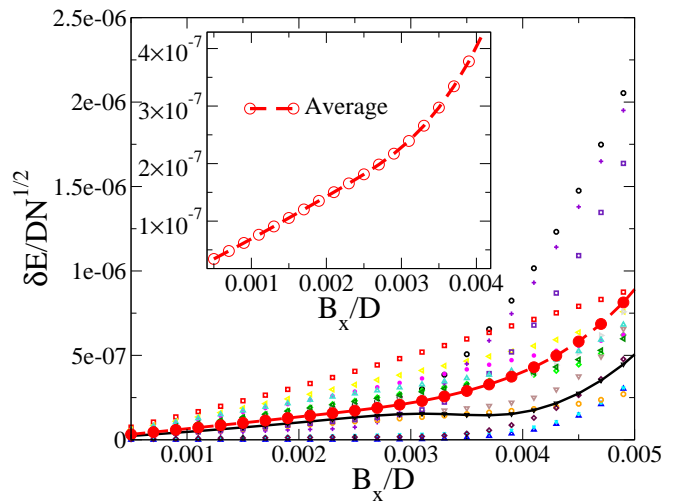}
\caption[figure3]{\label{fluctuations}(Color Online)
Random variations of the disorder configurations for a N=6-spin system for
twenty realizations of disorder.
Gap $\delta E/(D\sqrt{N})$ as a function of the transverse field $B_x/D$.
Depending on the disorder configuration the curves exhibit a local
maximum and a local minimum.
The thin (black) curve in the main panel shows the minimum/maximum
structure of $\delta E$ vs $B_x$ for a specific realization
of disorder. This structure disappear after taking the average as shown
by the thick (red) curve joining the filled (red) circles.
The monotonous behavior for the average of $\delta E$, already for 20 samples,
is emphasized in the inset.
}
\end{figure}

Interestingly, for a specific realization of disorder, in
Fig.~\ref{compare}, we note a local maximum in $\delta E$ around
$B_{x}/D\approx 0.0025$, followed by
a local minimum, before $\delta E$ starts diverging with increasing $B_x$.
We investigated the origin of this behavior and found that it can be
understood
as arising from the
$B_x$ dependence of $C_{zz}\propto (1-\frac{1}{2}(B_{x}/\Omega_0)^2)$
vs $C_{x0} \propto B_x/\Omega_0$,
both for small $B_x/D$.
Obviously, if this is the case, the random distribution of the magnetic ions
in the sample must play a crucial role in the position (and even the existence) of
this local maximum/minimum feature. The structure of $\delta E$ vs $B_x$
is controlled by the
$C_{\mu\nu}$ parameters, but not only: there is also a prefactor coming from
the dipolar interaction which is proportional to
$r_{ij}^z r_{ij}^x$. If one takes an extreme
case
in which
all
the magnetic ions are aligned on a line
along the $\hat z$ direction,
 the resultant interaction is 0,
and there is no dip in the curve. To confirm this scenario we show in
Fig.~\ref{fluctuations}  $\delta E/(D\sqrt{N})$ as a function of the
transverse field $B_x$ for twenty different disorder configurations for $N=6$.
One sees that the majority of curves do not show these
local maximum/minimum features and, as shown by the inset of Fig.~\ref{fluctuations},
the average of $\delta E$ over those twenty realizations of disorder reveal
no such max/min structure.

\begin{figure}[]
\includegraphics*[width=\columnwidth,angle=0]{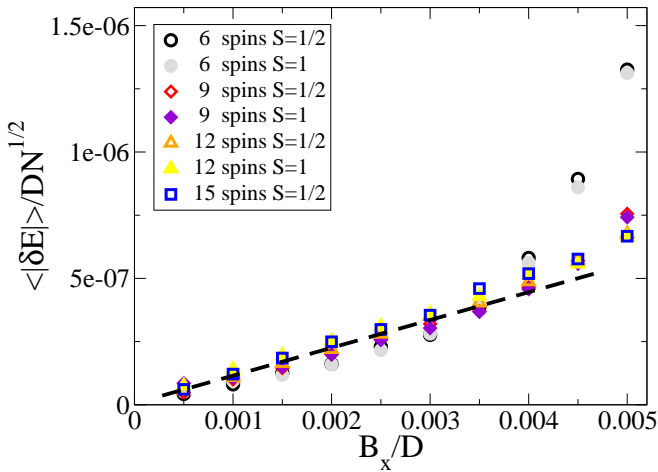}
\caption[figure3]{\label{average} (Color Online)
Scaling of the renormalized gap $\langle\delta E\rangle/\sqrt{N}$
(average taken over 1000 samples)
for various system sizes as a function of the transverse field $B_x$.
The closed symbols are for the $S=1$ model,
open symbols for the effective $S_{\rm eff}=\frac{1}{2}$ Hamiltonian.}
\end{figure}

Having demonstrated the one-to-one correspondence between the
$S=1$ and the effective $S_{\rm eff}=\frac{1}{2}$ model for
various (specific) realizations of disorer, we now proceed to
check the scaling with system size for $\langle \vert \delta
E\vert\rangle$ predicted by Eq.~(\ref{deltaEAn}) for the $S=1$
model and also check that it it agrees with the one for the
effective $S_{\rm eff}=\frac{1}{2}$ model The results for both
models are shown in Fig.~\ref{average}. The average gap $\langle
\vert \delta E \vert \rangle$ was computed over 1000 samples
which, for each system size of $N$ spins, we renormalize  as
$\langle \vert \delta E \vert \rangle/(D\sqrt{N})$, and plot for
both models ($S=1$ and $S_{\rm eff}=\frac{1}{2}$) as a function of
the transverse field $B_x$. As showed in
Ref.~[\onlinecite{Schechter-PRL2}] there exist a regime for which
the spin $S=1$ model obeys $\frac{\langle\vert \delta E \vert
\rangle}{\sqrt{N}}\propto \frac{B_x}{D}$ scaling. Indeed, for the
$S=1$ case (closed symbols), we clearly observe in
Fig.~\ref{average} a good collapse of the curves for the various
system sizes with
 this linear behavior.

One can see that at higher $B_x$, the scaling relation for different system
size $N$, as well as the proportionality of the gap
$\langle\vert \delta E \vert \rangle$
with $B_x$ starts to break down. As explained above in the context of
Fig.~\ref{compare},  this comes from the fact that the transverse
field term in the Hamiltonian is larger than the dipolar interaction
$V_{ij}^{zz}$. Thus the droplet picture is not valid and neither are the
scaling nor the proportionality relations in Eq.~(\ref{deq})
fulfilled. In Fig.~\ref{average}, we also show the
results for the effective $S_{\rm eff}=\frac{1}{2}$ model (open symbols),
demonstrating the agreement
with the results for the $S=1$ model,
even when
the  ($\delta E/\sqrt{N} \propto B_x/D$) regime breaks down.
This confirms the correctness of the conclusion based on Eq.~(\ref{cost}), and
that $\delta E$ is the same for both the $S=1$ and the $S_{\rm eff}=1/2$ models.

\section{Conclusion}
\label{Conclusion}

We have shown how to rigorously derive
an effective spin$-\frac{1}{2}$ Hamiltonian
to describe the problem of induced
random fields  in a  spin glass model
with strong single-ion Ising anisotropy and subject to a transverse magnetic field.
We discussed the relation of this problem with that of
the LiHo$_x$Y$_{1-x}$F$_4$ material in a magnetic field transverse to the
Ho$^{3+}$ Ising spins~\cite{Wu}.
We have shown, both analytically and numerically,
that the use of such a model
give results in full quantitative agreement with previously reported perturbation
theory calculations on a ``large'' spin $S$ model with strong
 anisotropy.\cite{Schechter-PRL2,Schechter-Osaka,Schechter-3}
However,
the large hyperfine interactions present in
the  {\it real} LiHo$_x$Y$_{1-x}$F$_4$, and which have been ignored here,
 must ultimately be considered
in order to obtain a good quantitative understanding of
the low-temperature regime.\cite{Schechter-PRL1}

The approach of Refs.~[\onlinecite{Schechter-PRL2,Schechter-Osaka}]
 proceeds
via the Rayleigh-Schr\"odinger perturbation theory, the one in
Ref.~[\onlinecite{Tabei}] and presented
in Section~\ref{Effective-H} above
relies on the effective Hamiltonian approach.
To low order in the quantum
${\cal H}_\perp$ term,
the two approaches have been shown to give  identical results.
However,  the emergence of induced random fields is much more apparent
in  the spin-$\frac{1}{2}$ effective model approach.
 The $C_{\mu\nu}$ coefficients needed to construct the effective Hamiltonian
are easily calculated,  providing an ability to
investigate the evolution of $\delta E$ and $\xi$ with $B_x$ beyond the linear term and to
arbitrarily high order in $B_x$. Such high order perturbation theory would be more cumbersome
to construct
when proceeding via a direct Rayleigh-Schr\"odinger perturbation scheme.
The crucial step connecting the perturbation theory method and the
effective $S_{\rm eff}=\frac{1}{2}$ approach is in the
determination of the $B_x$ dependence of the  $C_{\mu\nu}$ transformation parameters
in Eq.~(\ref{Cmunu}).
It is the neglect of this $B_x$ dependence of the spin interactions in
the $S_{\rm eff}=\frac{1}{2}$ model
investigated in Refs.~[\onlinecite{Schechter-Osaka,Schechter-3}]
that seemingly led their authors
to argue for the quantitative
 inadequacies of
the $S_{\rm eff}=\frac{1}{2}$  approach.

We note that, in a general case where the $V_{ij}^{zx}$
spin-spin interactions are not much smaller than $D$,
higher-order perturbation theory calculations must be
carried out to derive an effective Hamiltonian.
The physical result would be
that virtual transitions to the excited states lead
to an admixing of those states with the low-energy sector.
This effect  was recently discussed in Ref.~[\onlinecite{Chin}], where
it was shown that such interaction-induced quantum mechanical effects are
seemingly negligible for  LiHo$_x$Y$_{1-x}$F$_4$.
This makes difficult to understand the advocated phenomenon
of quantum mechanical entanglement proposed in Ref.~[\onlinecite{Ghosh-Nature}]
to explain the peculiar behavior of the
very dilute LiHo$_x$Y$_{1-x}$F$_4$ ($x=0.045$).
However, as a counter-example and for a different magnetic rare-earth system,
we note that it was recently found that
such interaction-induced admixing can dramatically change the low-energy physics.\cite{TTO}

With the contributions of
Refs.~[\onlinecite{Schechter-PRL1,Schechter-PRL2,Schechter-Osaka,Tabei,Schechter-3}]
and the clarification presented herein,
it may be that the behavior of {\it dilute}
LiHo$_x$Y$_{1-x}$F$_4$ in a transverse field, both in the random ferromagnetic and spin
glass regimes, are now somewhat understood. This impression would seem to be further
corroborated by recent experimental studies which provide evidence
for  the manifestation
of induced random fields for LiHo$_x$Y$_{1-x}$F$_4$
with $x=0.44$ and $B_x>0$. \cite{Silevitch-Nature}
Yet, there are many questions still opened regarding the physics of
this material for $x\lesssim 20\%$:
Is there a dipolar spin glass phase
over a reasonably wide range of dipole moment
concentration, either theoretically~\cite{Ghosh-Nature,Yu,Henelius}
or experimentally~\cite{Reich,Ghosh-Nature,Jonsson}?
What are the physical objects giving rise to the peculiar coherent dynamics at low-temperature
for samples with low Ho$^{3+}$ concentration
(see Refs.~[\onlinecite{Ghosh-Science,Silevitch-PRL}])?
Even for pure LiHoF$_4$, what is the microscopic explanation for the discrepancy
between experiment and Monte Carlo simulations
for the temperature vs transverse-field phase diagram for small $B_x$ near the classical
paramagnetic phase boundary? \cite{Chakraborty,Tabei-MC}
Are the phenomena found in zero and nonzero $B_x$ for LiHo$_x$Y$_{1-x}$F$_4$
also observable in other Ising systems which possess either Kramers or non-Kramers
rare-earth magnetic ions,\cite{Stasiak} or where the hyperfine interactions, so
important for Ho$^{3+}$ ions,\cite{Schechter-PRL1,Ramirez-Jensen,Bramwell}
may be much less significant~\cite{Stasiak}?
While it is interesting that LiHo$_x$Y$_{1-x}$F$_4$ in
a transverse field becomes a rare, if not the first physical realization of a random-field
Ising model in a ferromagnetic setting,\cite{Silevitch-Nature,Tabei,Schechter-3,Brooke-thesis}
it would seem that this is a small part of the challenges offered by this
material, with apparently more left to understand than has so far
been understood.


\section{Acknowledgments}

We thank Steve Girvin,
Helmut Katzgraber and Pawel Stasiak for useful discussions,
 and Nicolas Laflorencie for comments
on a previous version of this manuscript.
Support for this work was provided by NSERC
of Canada,  the Canada Research Chair Program (Tier I, M.G),
the Canada Foundation for Innovation,
the Ontario Innovation Trust, and the Canadian
Institute for Advanced research.
M.G. acknowledges the University of Canterbury (UC) for financial support
and the hospitality of the Department of Physics and Astronomy at UC
where part of this work was completed.


\begin{thebibliography}{99}
\bibitem{antiglass} An exotic ``antiglass'' state,
whose
nature is not yet well understood,
occurs
 at low concentration $x$ of Ho$^{3+}$ is
also observed. See Refs.~[\onlinecite{Ghosh-Science,Ghosh-Nature,Silevitch-PRL}].
\bibitem{B2s}
For S=1, using  second order degenerate perturbation theory,  one finds that
$\Delta=B_x^2\langle 1|S^x|0\rangle\langle 0|S^x|1\rangle/\Omega_0$.
For $S>1$, one should go to higher orders of perturbation
 theory obtaining the leading term proportional to $(B_x)^{2S}$.
\bibitem{up}
For $B_x=0$, $|\! \uparrow\rangle=|S\rangle$ and  $|\! \downarrow\rangle=|-S\rangle$,
which indicates that at $B_x=0$, $|\psi\rangle=|\Phi_S\rangle$.
\bibitem{S}
For S=1, $|\! \uparrow\rangle=\left(1-\frac{B_x^2}{4\Omega_0}\right) |1\rangle +
\frac{B_x}{\sqrt{2}\Omega_0^2}|0\rangle-\frac{B_x^2}{4\Omega_0^2}|-1\rangle$ and
$|\! \downarrow\rangle=\left(1-\frac{B_x^2}{4\Omega_0^2}\right) |-1\rangle +
\frac{B_x}{\sqrt{2}\Omega_0}|0\rangle-\frac{B_x^2}{4\Omega_0^2}|1\rangle$, but the final
 $B_x$ dependence of $C_{\mu\nu}$'s are, to leading order,  not changed, and are similar to the
$S>1$ case.
\bibitem{boundary}
Since our goal is merely to compare numerically the energy eigenvalues
of two quantum spin models, as opposed to describe a system approaching the
thermodynamic limit, we consider here the simplest case of open boundary conditions
as opposed to periodic boundary conditions, which could be implemented,
for example, using Ewald summation techniques as done in Refs~[\onlinecite{Stasiak,Tabei-MC}].
\bibitem{droplet}
This regime corresponds to a region where the transverse field $B_x$ is
small compared to the dipolar intercation $V^{zz}$
\end{thebibliography}
\end{document}